\documentclass{PoS}

\usepackage{amssymb}  
\usepackage{amsthm}
\usepackage{amsmath}
\usepackage{siunitx}
\usepackage{url}
\usepackage{longtable}
\usepackage{graphicx}
\usepackage{verbatim}

\title{$B_{(s)}\to D_{(s)}$ semileptonic decays with NRQCD-HISQ valence quarks}

\ShortTitle{$B_{(s)}\to D_{(s)}$ semileptonic decays}

\author{\speaker{Christopher J.~Monahan}\\
New High Energy Theory Center and Department of Physics and 
Astronomy, Rutgers, The State University of New Jersey,
136 Frelinghuysen Road, Piscataway, NJ 08854, USA\\
E-mail: \email{chris.monahan@rutgers.edu}}

\author{Heechang Na\\
Ohio Supercomputer Center, 1224 Kinnear Road, Columbus, OH 
43212, USA\\
Department of Physics and Astronomy, 
University of Utah, 
Salt Lake City, UT 84112, USA
}
\author{Chris M.~Bouchard\\
School of Physics and 
Astronomy, University of Glasgow, 
Glasgow G12 8QQ, UK\\
Department of Physics and Astronomy, College of 
William and Mary, Williamsburg, VA 23187, USA
}
\author{{G.~Peter} Lepage\\
Laboratory of Elementary Particle Physics,
Cornell University, Ithaca, NY 14853, USA}
\author{Junko Shigemitsu\\
Department of Physics,
The Ohio State University, Columbus, OH 43210, USA}

\abstract{We present a calculation of the form factors, $f_0$ and $f_+$, for 
the $B_{(s)} \to D_{(s)}$ semileptonic decays. Our work uses the MILC $n_f=2+1$ 
AsqTad configurations with NRQCD and HISQ valence quarks at four values of the 
momentum transfer $q^2$. We provide results for the
chiral-continuum extrapolations of the scalar and vector form factors.}

\FullConference{34th annual International Symposium on Lattice Field Theory\\
		24-30 July 2016\\
		University of Southampton, UK}

\begin{document}

\section{Introduction}

Precision measurements of $B$ and $B_s$ meson decays at the Large Hadron 
Collider are an 
important tool in the search for new physics. For example, the first 
observation of 
the rare decay $B_s \rightarrow \mu^+ 
\mu^-$, through a combined analysis by the LHCb and CMS collaborations 
\cite{CMS:2014xfa}, tested the Standard Model prediction of the 
branching fraction. In the Standard Model this decay is doubly-suppressed, but 
the branching fraction may receive large contributions from new physics 
effects. Currently, the measured branching fraction is consistent with 
Standard Model expectations, but there is still room for 
new physics, given the experimental and theoretical 
uncertainties. Run II at the LHC should significantly reduce experimental 
uncertainties for a wide range of $B_{(s)}$ decays. Tightening
constraints on potential new physics therefore requires a similar
improvement in the theoretical determination of the Standard Model expectations.

The $B_s$ meson branching fraction 
${\cal B}( B_s \rightarrow \mu^+ \mu^-)$ can be expressed in terms of the ratio 
of fragmentation fractions, $f_d/f_s$. The fragmentation fraction $f_q$ gives 
the 
probability that a $b$-quark hadronizes into a $B_q$ meson. Reducing 
sources of systematic uncertainties in the value of this ratio will 
improve not just the precision of the determination of the $B_s \rightarrow 
\mu^+ \mu^-$ 
branching fraction, but a range of other $B_s$ 
meson decay branching fractions at the LHC as well \cite{Adeva:2009ny}.

The ratio of the fragmentation fractions, $f_d/f_s$, can be extracted from the 
ratio of the scalar form factors of the $B \rightarrow D l \nu$ and
$B_s \rightarrow D_s l \nu$ semileptonic decays 
\cite{Fleischer:2010ay}.
There is currently only one lattice determination of the form factor 
ratio, using heavy clover bottom and charm quarks \cite{Bailey:2012rr}. The 
form 
factors, $f_+(q^2)$ and $f_0(q^2)$, for 
the semileptonic decay $B_s \rightarrow D_s \ell \nu$ were determined with 
twisted mass fermions for the region near zero recoil in \cite{Atoui:2013zza}.

In addition to determining the fragmentation ratio relevant to the 
measurement of the branching fraction for the rare decay, $B_s \rightarrow 
\mu^+ \mu^-$, the semileptonic $B_s\to D_s\ell\nu$ decay provides a new method 
to determine the CKM matrix element $|V_{cb}|$. There is a 
long-standing tension between determinations of $|V_{cb}|$ from exclusive 
and inclusive measurements of the 
semileptonic $B$ meson decays (see, for example, 
\cite{pdg15}), although recent analyses suggest the tension has
eased \cite{Gambino:2016jkc}. Future experimental observation 
of the $B_s\to D_s\ell\nu$ decay, combined with lattice predictions of the form 
factors, may shed light on the $V_{cb}$ puzzle.

We report on the HPQCD collaboration's calculations of the form 
factors, $f_+(q^2)$ and $f_0(q^2)$, for 
the semileptonic decays $B_{(s)} \rightarrow D_{(s)} \ell \nu$, using the 
non-relativistic (NRQCD) action for the bottom quarks and the Highly 
Improved Staggered Quark (HISQ) action for the charm quarks. 
Our results for the $B \rightarrow D \ell \nu$ decay appeared first
in \cite{Na:2015kha}.  We refer the reader to Sections II and III of 
\cite{Na:2015kha} for further details of the analysis.

\section{\label{sec:lattsetup}Ensemble details}

We use five gauge ensembles, summarized in Table \ref{tab:milc}, generated by 
the MILC collaboration \cite{Bazavov:2009bb}. These ensembles include three 
``coarse'' (with 
lattice spacing $a \approx \SI{0.12}{fm}$) and two ``fine'' (with $a \approx 
\SI{0.09}{fm}$) ensembles, incorporating $n_f = 2+1$ flavors of AsqTad sea 
quarks.
\begin{table}
\caption{\label{tab:milc}
Simulation details on three ``coarse'' and two ``fine''  $n_f = 2 + 1$ MILC 
ensembles.
}

\begin{tabular}{cccccccccc}
Set &  $r_1/a$ & $m_l/m_s$ (sea)   &  $N_{\mathrm{conf}}$&
$N_{\mathrm{tsrc}}$ & $L^3 \times N_t$ & $a m_b$   & $a m_s$ & $a m_c$ &  
$aE_{b\overline{b}}^{\mathrm{sim}}$  \\
\vspace*{-10pt}\\
\hline 
\vspace*{-10pt}\\
C1  & 2.647 & 0.005/0.050   & 2096  &  4 & $24^3 \times 64$ & 2.650 & 0.0489 & 
0.6207 & 0.28356(15) \\
C2  & 2.618 & 0.010/0.050  & 2256   & 2 & $20^3 \times 64$& 2.688 & 0.0492 & 
0.6300 & 0.28323(18) \\
C3  & 2.644 & 0.020/0.050  & 1200  & 2 & $20^3 \times 64$& 2.650 & 0.0491 & 
0.6235 & 0.27897(20) \\
F1  & 3.699 & 0.0062/0.031  & 1896  & 4  & $28^3 \times 96$ & 1.832 & 0.0337 & 
0.4130 & 0.25653(14) \\
F2  & 3.712 & 0.0124/0.031  & 1200  & 4 & $28^3 \times 96$ & 1.826 & 0.0336 & 
0.4120 & 0.25558(28) \\
\end{tabular}
%\end{center}
\end{table}

We study $B_{(s)} \rightarrow D_{(s)}$ semileptonic decays by evaluating the 
matrix element of the bottom-charm vector current, $V^\mu$, between $B_{(s)}$ 
and 
$D_{(s)}$ states. We express these matrix elements in terms of the form factors
$f_+^{(s)}(q^2)$ and $f_0^{(s)}(q^2)$ as
\begin{equation}
\langle D_{(s)}(p_{D_{(s)}}) | V^\mu | B_{(s)}(p_{B_{(s)}}) \rangle = 
f_0^{(s)}(q^2) 
\frac{M_{B_{(s)}}^2-M_{D_{(s)}}^2}{q^2}q^\mu  +f_+^{(s)}(q^2) \left[
p_{B_{(s)}}^\mu + p_{D_{(s)}}^\mu - \frac{M_{B_{(s)}}^2-M_{D_{(s)}}^2}{q^2}q^\mu
\right],
\end{equation}
where the momentum transfer is $q^\mu = p_{B_{(s)}}^\mu - p_{D_{(s)}}^\mu$. 
In 
practice 
it is simpler to work with the form factors $f_\parallel^{(s)}$ and 
$f_\perp^{(s)}$, which are related to $f_+^{(s)}(q^2)$ and $f_0^{(s)}(q^2)$ via
\begin{align}
f_+^{(s)}(q^2) = {} & \frac{1}{\sqrt{2M_{B_{(s)}}}}\left[f_\parallel^{(s)}(q^2) 
+ 
(M_{B_{(s)}}-E_{D_{(s)}})f_\perp^{(s)}(q^2)\right], \\
f_0^{(s)}(q^2) = {} & 
\frac{\sqrt{2M_{B_{(s)}}}}{M_{B_{(s)}}^2-M_{D_{(s)}}^2}\bigg[(M_{B_{(s)}}-E_{D_{
(s)} }
)f_\parallel^{(s)}(q^2) + 
(E_{D_{(s)}}^2-M_{D_{(s)}}^2)f_\perp^{(s)}(q^2)\bigg].
\end{align}
Here $E_{D_{(s)}}$ is the energy of the daughter $D_{(s)}$ meson in the rest 
frame of 
the $B_{(s)}$ meson.

We calculate $B_{(s)}$ and $D_{(s)}$ meson two-point correlators and 
three-point 
correlators of the bottom-charm currents, $J_\mu$. We use smeared 
heavy-light or heavy-strange 
bilinears to represent the $B_{(s)}$ meson, with either delta-function 
or Gaussian smearing. Three-point correlators are computed 
as follows: The $B_{(s)}$ meson is 
created 
at time $t_0$; a current $J_\mu$ inserted at timeslice $t$; and the $D_{(s)}$ 
meson annihilated at timeslice 
$t_0+T$, where $t_0< t < t_0+T$. We use four values of $T$ and generate data 
for four values of the $D_{(s)}$ 
meson momenta, $\vec{p}_{D_{(s)}} \in 
2\pi/(aL)\{(0,0,0),(1,0,0),(1,1,0),(1,1,1)\}$, where $L$ is the spatial 
lattice extent. We work in the rest frame of the $B_{(s)}$ meson. 

We match the NRQCD-HISQ currents, $J_\mu$, at one loop in 
perturbation theory, that is, through ${\cal O}(\alpha_s, 
\Lambda_{\mathrm{QCD}}/m_b, \alpha_s/am_b)$, where 
$am_b$ is the bare lattice mass \cite{Monahan:2012dq}. We re-scale all currents 
by the tree-level massive wave function renormalization for the HISQ charm 
quarks \cite{Na:2015kha}.

\section{\label{sec:corrfits}Correlator fits}

We fit $B_{(s)}$ and $D_{(s)}$ meson two-point functions to a sum of decaying 
exponentials in Euclidean time, $t$,
\begin{align}
C_{B_{(s)}}^{\beta,\alpha}(t) = {} & \sum_{i=0}^{N_{B_{(s)}}-1} 
b_i^\beta  
b_i^{\alpha\ast} e^{-E_i^{B_{(s)},\mathrm{sim}}t}  + 
\sum_{i=0}^{N_{B_{(s)}}'-1}b_i^{\prime\,\beta} 
b_i^{\prime\,\alpha\ast} 
(-1)^te^{-E_i^{\prime\,B_{(s)},\mathrm{sim}}t}, \\
C_{D_{(s)}}(t;\vec{p}) = {} & \sum_{i=0}^{N_{D_{(s)}}-1} 
|d_i|^2 \left[e^{-E_i^{D_{(s)}}t}+e^{-E_i^{D_{(s)}}(N_t-t)}\right] 
 +\sum_{i=0}^{N_{D_{(s)}}'-1}|d_i'|^2 
(-1)^t\left[e^{-E_i^{\prime\,D_{(s)}}t}+e^{-E_i^{\prime\,D_{(s)}}(N_t-t)}\right]
\end{align}
The superscripts $\alpha$ and $\beta$ indicate the two forms of smearing for 
the $B_{(s)}$ meson source (delta function or 
Gaussian). The amplitudes 
associated with the ordinary and oscillatory states are $b_i$ and 
$b_i'$, with associated meson energies
$E_i^{B_{(s)},\mathrm{sim}}$ and $E_i^{\prime\,B_{(s)},\mathrm{sim}}$, and 
$d_i$ and $d_i'$, where the corresponding meson energies are
$E_i^{D_{(s)}}$ and $E_i^{\prime\,D_{(s)}}$, 
respectively. The ground state $B_{(s)}$ energy 
in NRQCD, $E_0^{B_{(s)},\mathrm{sim}}$, is not equal to 
the true energy in full QCD, $E_0^{B_{(s)}}$,
because the $b$-quark rest mass has been integrated out in NRQCD. Here 
$\overline{M}_{b\overline{b}}^{\mathrm{exp}}$ is the spin-averaged $\Upsilon$ 
mass used to tune the $b$-quark mass and $aE_{b\overline{b}}^{\mathrm{sim}}$ 
was 
determined in \cite{Na:2012kp}. For both $B_{(s)}$ and $D_{(s)}$ two-point 
functions, $N$ is the number of exponentials included in the fit.

For the three-point correlator, we use an ansatz that incorporates four terms, 
each of which is a sum of exponentials, similar to the two-point forms shown 
above. For the full form, see \cite{Na:2015kha}. 
We then determine the hadronic matrix element between $B_{(s)}$ and $D_{(s)}$, 
in the rest frame of the $B_{(s)}$ meson, from
\begin{equation}
\langle D_{(s)}(p) | V^\mu | B_{(s)} \rangle = 
\frac{A_{00}^\alpha}{d_0b_0^{\alpha\ast}}\sqrt{2a^3E_0^{D_{(s)}}}\sqrt{2a^3M_{
B_{(s)}}}.
\end{equation}

We fit these correlators using Bayesian multi-exponential fitting 
\verb+python+ packages \verb+lsqfit+ and \verb+corrfitter+ \cite{lsqfit}, an 
approach that has been 
used by the HPQCD collaboration for a wide range of lattice calculations. 

We tested a number of indicators of 
fit stability, consistency, and goodness-of-fit to check the fit 
results. For example, we checked that, beyond a minimum number of exponentials, 
the fit results are independent of the number of exponentials included in the 
fit. We tested three types of fits: simultaneous fits to correlator 
data for all four spatial momenta; chained fits 
(discussed in detail in the appendices of \cite{Bouchard:2014ypa}) to 
correlator data for all four spatial momenta simultaneously; and ``individual'' 
fits, including the correlator data for just a single daughter 
meson momentum in each fit. All three fit approaches give consistent results.

The simultaneous fits, with or 
without chaining, have the advantage that they capture the correlations between 
momenta, which is then reflected in the uncertainty quoted in the fit results. 
The chained fits give slightly better values of reduced $\chi^2$ and are about 
ten percent faster than the 
simultaneous 
fits, which is an important consideration for the large three-point fits.  
We take the 
result for $N_{\mathrm{exp}} = 5$ from the chained fit as our final 
result for each momentum.  Choosing to use chained fits for 
both two- and three-point fits ensures a consistent approach throughout the 
fitting procedure.

\begin{figure}
\centering
\caption{\label{fig:dsdispersion}Dispersion relation
for the ensemble sets C2 and F1. For the $B_s\to D_s\ell\nu$ decay (right-hand 
pane), we 
include all three types of fits listed in the 
text (simultaneous, chained, and individual) to illustrate the consistency of 
our results. The shaded region 
corresponds to $1\pm\alpha_s(ap/\pi)^2$ where we take $\alpha_s=0.25$. The 
left-hand pane is taken from \cite{Na:2015kha}.}
\includegraphics[width=0.45\textwidth,keepaspectratio]{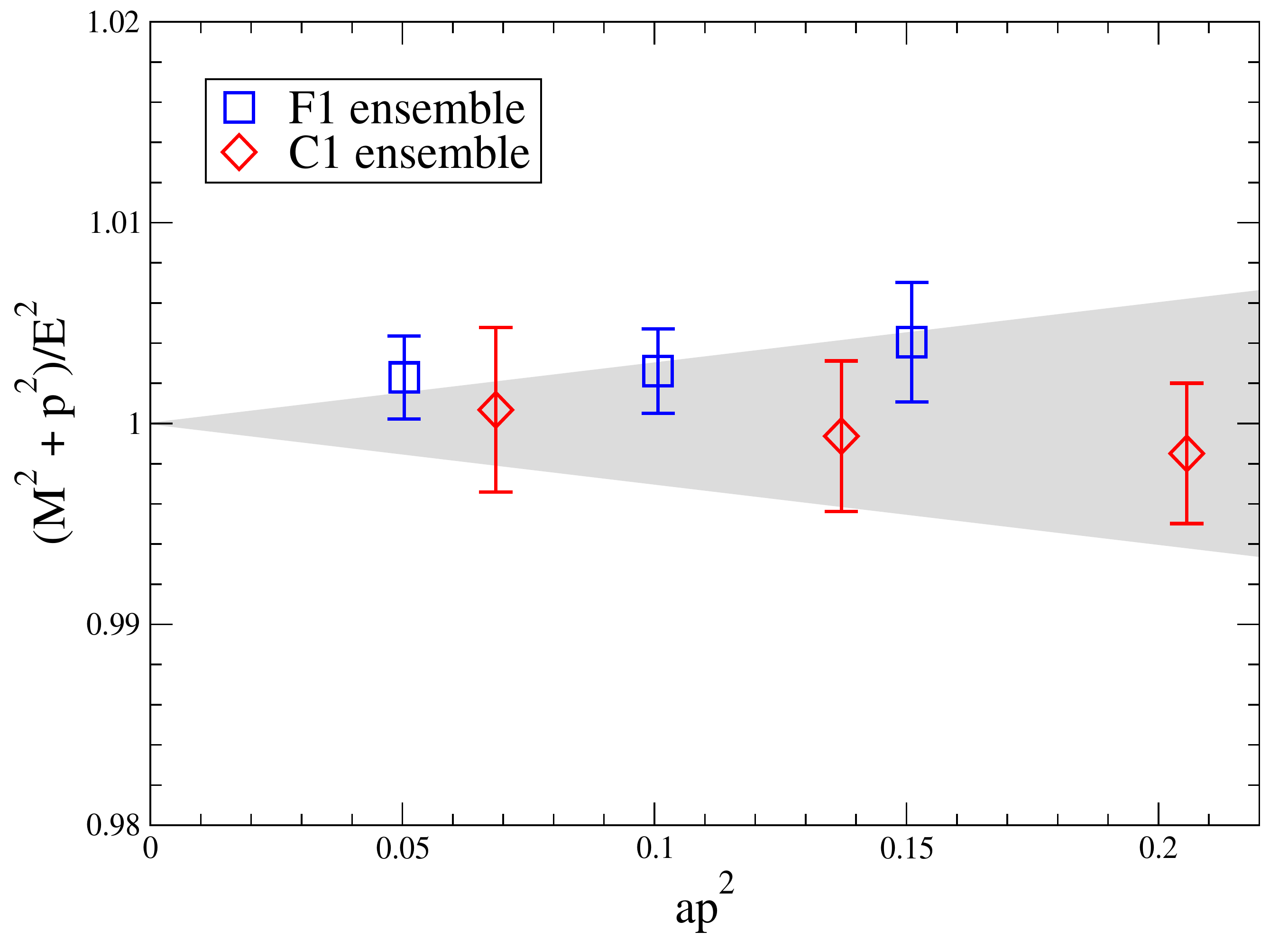}
\includegraphics[width=0.48\textwidth,keepaspectratio]{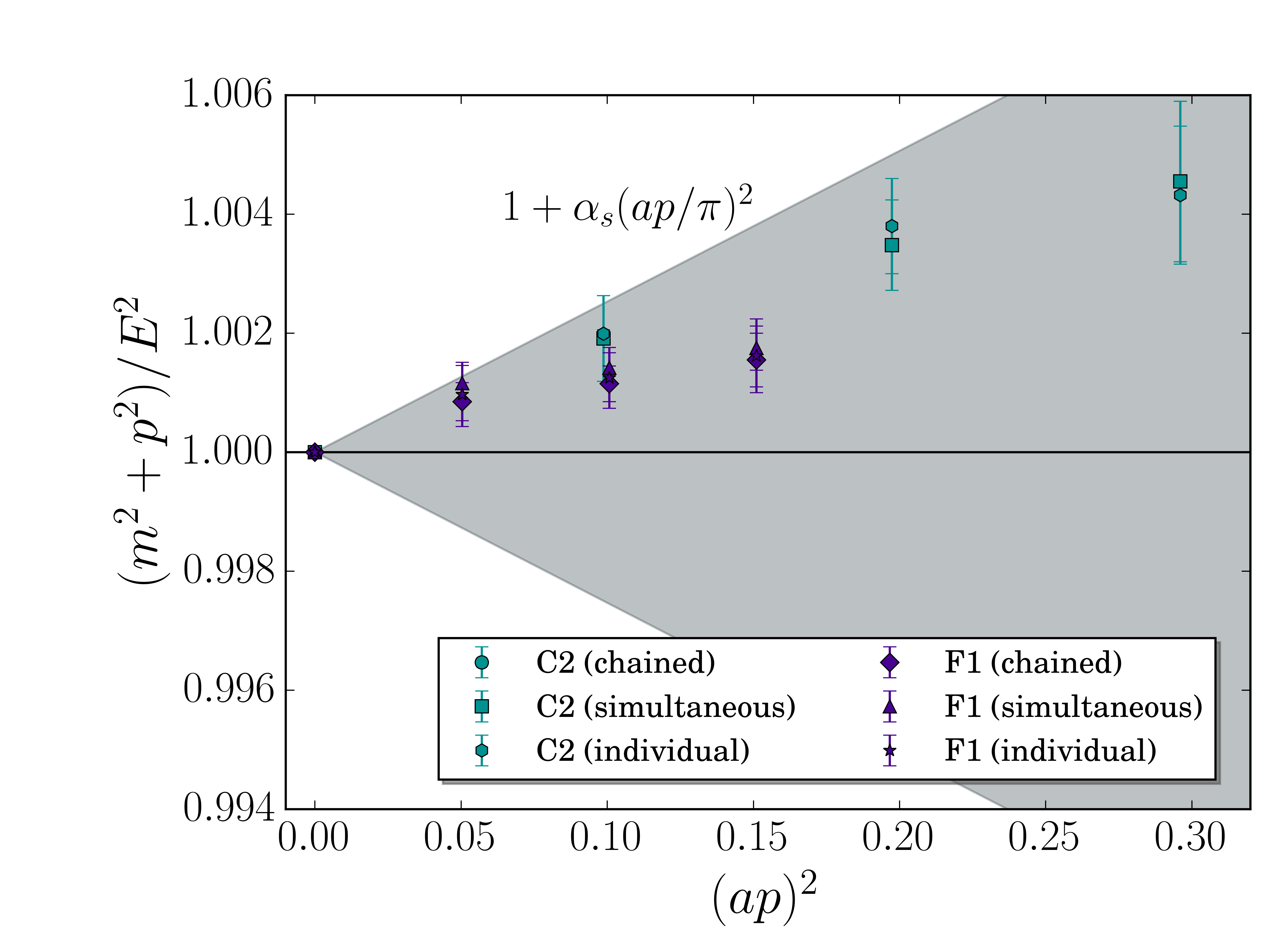}
\end{figure}

For each ensemble, we determined the 
ratio $(M_{D_{(s)}}^2+\vec{p}^2)/E_{D_{(s)}}^2$ and illustrate some results 
in Figure \ref{fig:dsdispersion}. The shaded region corresponds to $1\pm 
\alpha_s(ap/\pi)^2$, where we set $\alpha_s = 0.25$. In contrast to the 
$B \to D \ell\nu$ case, the data for the $B_s \to D_s\ell \nu$ decay lie 
systematically above the relativistic value of unity, indicating that the 
statistical uncertainties of the fit results are sufficiently small that we 
can resolve discretization effects at ${\cal O}(\alpha_s(ap/\pi)^2)$. 
In both cases, the discretization effects are less than $0.5\%$ in the 
dispersion relation.

We summarize our preliminary results for the form factors, $f_0^{(s)}(\vec{p})$ 
and 
$f_+^{(s)}(\vec{p})$, for each ensemble and $D_{(s)}$ momentum, in Table
\ref{tab:f0fp}. 
\begin{table}
\caption{\label{tab:f0fp}
Preliminary results for the form factors, $f_0^{(s)}(\vec{p})$ and 
$f_+^{(s)}(\vec{p})$. For the $B\to D\ell \nu$ decay, the results are taken 
from \cite{Na:2015kha}. 
From top to bottom, the rows correspond to ensemble sets C1, C2, C3, F1, and 
F2.}\vspace*{\baselineskip}
\begin{tabular}{ccccccc}
$f_0(0,0,0)$   & $f_0(1,0,0)$ & $f_0(1,1,0)$ & $  
f_0(1,1,1)$ & $f_+(1,0,0)$   & $f_+(1,1,0)$ & $f_+(1,1,1)$ \\
\vspace*{-10pt}\\
\hline 
\vspace*{-8pt}\\
0.8810(56) & 0.8743(43)& 0.8608(38)& 0.8534(42) & 1.135(12) &  1.1125(57) &  
1.0837(61) \\
0.8809(31) & 0.8716(54) & 0.8617(44) & 0.8503(50) & 1.110(12) & 1.0809(70) & 
1.0479(64) \\
0.8872(23) & 0.8685(32) & 0.8592(29) & 0.8473(38) & 1.1282(71) &  1.0937(40) &  
1.0569(50) \\
0.9034(31) & 0.8771(42) &0.8643(41) &0.8479(56) & 1.1344(91) &  1.0931(59)  &  
1.0480(74) \\
0.9051(23) &0.8895(36) &0.8702(29) &0.8504(34) &  1.1461(72) & 1.0963(39) &  
1.0577(45)\\
\\
$f_0^s(0,0,0)$   & $f_0^s(1,0,0)$ & $f_0^s(1,1,0)$ & $  
f_0^s(1,1,1)$ & $f_+^s(1,0,0)$   & $f_+^s(1,1,0)$ & $f_+^s(1,1,1)$ \\
\vspace*{-10pt}\\
\hline 
\vspace*{-8pt}\\
0.8885(11) & 0.8754(14) & 0.8645(13) & 0.8568(13) & 1.1384(35) & 1.1081(20) 
& 1.0827(21)\\
0.8822(13) & 0.8663(15) & 0.8524(16) & 0.8418(18) & 1.1137(29) & 1.0795(22) 
& 1.0470(21)\\
0.8883(13) & 0.8723(16) & 0.8603(16) & 0.8484(21) & 1.1260(34) & 1.0912(24) 
& 1.0552(28)\\
0.9063(10) & 0.8848(13) & 0.8674(13) & 0.8506(17)  & 1.1453(29) & 
1.0955(24) & 1.0549(24)\\
0.9047(12) & 0.8855(16) & 0.8667(15) & 0.8487(19) & 1.1347(42) & 1.0905(26) 
& 1.0457(33)
\end{tabular}
\end{table}

\section{\label{sec:zexp}Chiral, continuum and kinematic extrapolations}

We express the dependence of the form factors on the $z$-variable, 
$z(q^2) = \frac{\sqrt{t_+-q^2} - \sqrt{t_+-t_0}}{\sqrt{t_+-q^2}+\sqrt{t_+-t_0}}
$, where $t_+ = (M_{B_{(s)}}+M_{D_{(s)}})^2$ and we take $t_0 = 
q_{\mathrm{max}}^2$, through a 
modification of the BCL 
parameterization \cite{Bourrely:2008za}
\begin{align}
f_0^{(s)}(q^2(z)) = {} & \frac{1}{P_0}\sum_{j=0}^{J-1} 
a_j^{(0,(s))}(m_l,m_l^{\mathrm{sea}},a) 
z^j, \\
f_+^{(s)}(q^2(z)) = {} &  \frac{1}{P_+}\sum_{j=0}^{J-1} 
a_j^{(+,(s))}(m_l,m_l^{\mathrm{sea}},a)
\left[z^j - (-1)^{j-J}\frac{j}{J}z^J\right].
\end{align}
Here the $P_{0,+}$ are Blaschke factors that take into account the effects of 
expected poles above the physical region. The expansion coefficients 
$a_j^{(0,+,(s))}$ include 
lattice spacing and light quark mass dependence. 
We modify this parameterization of the form factors to 
accommodate the systematic uncertainty associated with the truncation of the 
matching procedure at ${\cal O}(\alpha_s, \Lambda_{\mathrm{QCD}}/m_b, 
\alpha_s/(am_b))$. We introduce fit parameters $m_\parallel$ and $m_\perp$, 
with central value zero and width $\delta m_{\parallel,\perp}$ and re-scale the 
form factors, $f_\parallel$ and $f_\perp$ according to$
f_{\parallel,\perp} \rightarrow (1+m_{\parallel,\perp})f_{\parallel,\perp}$.
We take the systematic uncertainties in these fit parameters as 
3\% and refer the reader to the detailed discussion of this approach in 
\cite{Na:2015kha}.

\begin{figure}
\centering
\caption{\label{fig:ffres1}Fit results as a function of the momentum transfer, 
$q^2$, for the $B \to 
D\ell\nu$ (left) and $B_s \to 
D_s\ell\nu$ (right) semileptonic decays. The left-hand pane first appeared, in 
slightly modified form, in \cite{Na:2015kha}.}
\includegraphics[width=0.48\textwidth,keepaspectratio]{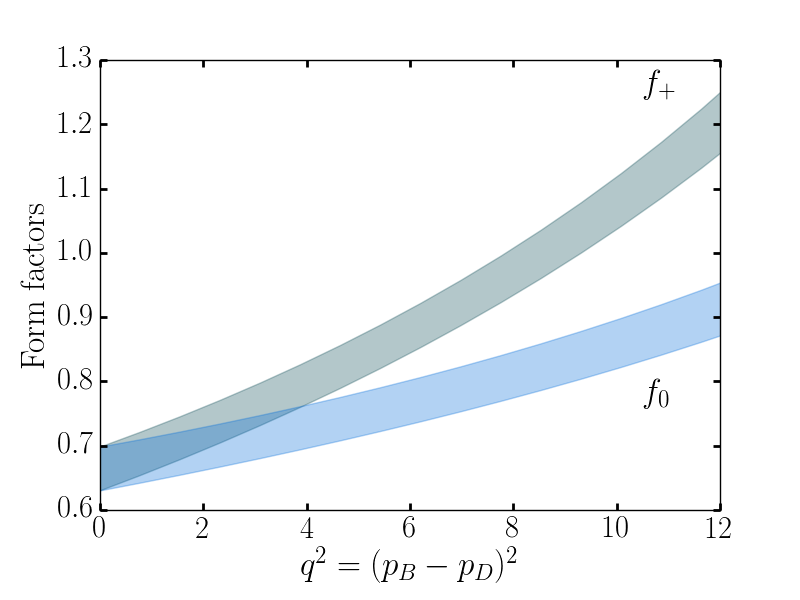}
\includegraphics[width=0.48\textwidth,keepaspectratio]{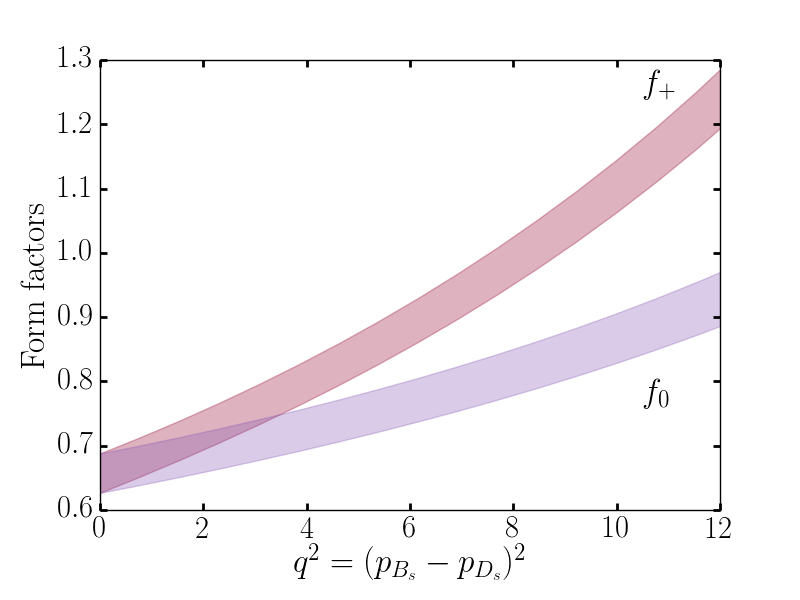}
\end{figure}

In Figure \ref{fig:ffres1}
we plot our fit results for 
$f_0^{(s)}(z(q^2))$, $f_+^{(s)}(z(q^2))$ as a function of the momentum 
transfer, $q^2$, 
for $B \to D\ell\nu$ (left panel) and $B_s \to D_s\ell\nu$ (right 
panel) semileptonic decays. We test the convergence of our fit ans\"atze by 
modifying the fit function, as outlined in detail in 
\cite{Na:2015kha}

\section{\label{sec:summary}Summary}

We have presented lattice calculations of the $B_{(s)} \to 
D_{(s)}\ell\nu$ semileptonic decays and determined
the form factors, $f_0^{(s)}(q^2)$ and $f_+^{(s)}(q^2)$ over the full kinematic 
range of momentum transfer. There are currently a number of tensions between 
experimental measurements and 
theoretical expectations for semileptonic decays of the $B$ 
meson. These tensions include the branching fraction ratios, $R(D^{(\ast)})$, 
and determinations of $|V_{cb}|$ from exclusive and inclusive decays. Future 
experimental measurements of semileptonic decays of $B_{s}$ mesons, in 
conjunction with our results for the form factors, may 
provide some insight into these tensions.

The dominant uncertainties in the form factors for the $B_{(s)} \to 
D_{(s)}\ell\nu$ decays arise from the discretization effects, with a 
significant 
contribution from the matching to full QCD. Higher order calculations in 
lattice perturbation theory with the highly 
improved actions used in this calculation are currently unfeasible, so we 
are exploring methods to reduce matching errors by combining results 
calculated using NRQCD with those determined with an entirely relativistic 
formulation for the $b$-quark 
\cite{Na:2015kha,Bouchard:2014ypa}.

The LHC is scheduled to significantly improve the statistical 
uncertainties in experimental measurements of $B_{(s)}$ decays with more data 
over 
the next decade. Reduced uncertainties on the corresponding form factors will 
improve theory errors in the fragmentation function 
ratio, 
$f_s/f_d$, used to extract branching fractions of $B_s$ decays 
at the LHC, and in determinations of $|V_{cb}|$. These improvements 
will be necessary to exploit fully the improved statistical 
precision of future experimental results and ultimately shed light on current
tensions in the heavy quark flavor sector.

\begin{acknowledgments}
Numerical simulations were carried out on facilities
of the USQCD collaboration funded by the Office of Science of the DOE and 
at the Ohio Supercomputer Center. This work was supported in part by grants 
from the 
DOE and NSF.
We thank the MILC collaboration for use of their gauge configurations.
\end{acknowledgments}

\end{document}